\documentclass[10pt, final, journal, letterpaper, twocolumn]{IEEEtran}
\usepackage{amsmath}
\usepackage{amssymb}
\usepackage{amsfonts}
\usepackage{caption}
\usepackage{graphicx}
\usepackage{epsfig}
\usepackage{epsf}
\usepackage{subfigure}
\usepackage{psfrag}
\usepackage{cite}
\usepackage{latexsym}
\usepackage{url}
\usepackage{color}
\usepackage{bm}
\usepackage{cases}
\usepackage{empheq}
\usepackage{algorithm}
\usepackage{algpseudocode}
\usepackage{graphics}
\usepackage{multirow}

\newcommand{\mv}[1]{\mbox{\boldmath{$ #1 $}}}

\begin{document}
\captionsetup[figure]{name={Fig.},labelsep=period}
\title{Secure UAV Communication with Cooperative Jamming and Trajectory Control}
\author{\IEEEauthorblockN{Canhui~Zhong, Jianping~Yao, and Jie~Xu}
\thanks
{
The authors are with the School of Information Engineering, Guangdong University of Technology, Guangzhou, China (e-mails: zhongcanhui93@qq.com, yaojp@gdut.edu.cn, jiexu@gdut.edu.cn).
}
}
\maketitle

\begin{abstract}
 This paper presents a new cooperative jamming approach to secure the unmanned aerial vehicle (UAV) communication by leveraging jamming from other nearby UAVs to defend against the eavesdropping. In particular, we consider a two-UAV scenario when one UAV transmitter delivers the confidential information to a ground node (GN), and the other UAV jammer cooperatively sends artificial noise (AN) to confuse the ground eavesdropper for protecting the confidentiality of the data transmission. By exploiting the fully-controllable mobility, the two UAVs can adaptively adjust their locations over time (a.k.a. trajectories) to facilitate the secure communication and cooperative jamming. We assume that the two UAVs perfectly know the GN's location and partially know the eavesdropper's location {\emph{a-priori}}. Under this setup, we maximize the average secrecy rate from the UAV transmitter to the GN over one particular time period, by optimizing the UAVs' trajectories, jointly with their communicating/jamming power allocations. Although the formulated problem is non-convex, we propose an efficient solution by applying the techniques of alternating optimization and successive convex approximation (SCA).
\end{abstract}
\begin{IEEEkeywords}
Secure UAV communication, cooperative jamming, trajectory design, power allocation.
\end{IEEEkeywords}

\section{Introduction}
Unmanned aerial vehicle (UAV) communications have attracted growing research interests in both academia and industry, in order to provide UAVs with not only low-latency and reliable command and control, but also application-specific high-speed data transmission \cite{chanllenges,WuCapacity}. However, due to the strong line-of-sight (LoS) air-to-ground wireless links, UAV communications face more stringent security issues than conventional terrestrial communication systems, as UAVs' transmit signals are more likely to be overheard by suspicious eavesdroppers over a large area on the ground.
Therefore, how to ensure the confidentiality of UAV communications against malicious eavesdropping attacks is a challenging task to be tackled.

Recently, physical-layer security has emerged as a promising solution to secure UAV communications against eavesdropping attacks (e.g., \cite{zhangguangchi,cui,zhou,li,qian,zhu,zhangshuhang}).
For example, \cite{zhangguangchi,cui} studied a secure single-UAV communication system with one UAV communicating with a ground node (GN) in the presence of suspicious eavesdroppers on the ground.
The authors proposed to adaptively control the UAV's location over time (a.k.a. trajectory) jointly with the transmit power allocation to maximize the UAV's average secrecy rate over a finite mission period.
By exploiting the trajectory design, the UAV can fly towards the intended GN to enhance the communication quality and move far apart from the eavesdroppers to reduce the information leakage, thus significantly improving the secrecy rate.
Furthermore, \cite{zhou,li} proposed to use UAVs as friendly jammers to secure the ground wireless communication, while \cite{qian,zhangshuhang} considered to employ UAVs as mobile relays to facilitate secure or reliable wireless communications.
In addition, the authors in \cite{zhu} analyzed the secrecy performance of a large UAV-enabled millimeter wave (mmWave) network, where a large number of randomly deployed UAVs act as flying BSs to serve users on the ground.

Different from prior works, in this paper we propose a new inter-UAV cooperative jamming approach to secure the UAV communication, by allowing other nearby UAVs to cooperatively send jamming signals for confusing potential eavesdroppers. For ease of exposition, we consider a basic two-UAV scenario when one UAV transmitter sends information to a GN and the other UAV jammer generates artificial noise (AN) to jam a suspicious eavesdropper on the ground. By exploiting the fully-controllable mobility, the two UAVs can adaptively adjust their trajectories, together with the wireless resource allocation, to facilitate the secure communication and cooperative jamming. It is assumed that the UAVs perfectly know the GN's location and partially know the eavesdropper's location {\emph{a-priori}}.
Under this setup, we maximize the average secrecy rate from the UAV transmitter to the GN by optimizing the two UAVs' trajectories, jointly with the communicating/jamming power allocations over time.
Although this problem is non-convex and generally difficult to solve, we obtain a high-quality solution efficiently via employing the techniques of alternating optimization and successive convex approximation (SCA).
Numerical results show that the proposed design significantly improves the secrecy rate performance, as compared to other benchmark schemes.

Note that the proposed cooperative jamming design for secure UAV communications is different from that in conventional secure terrestrial communication systems (see, e.g., \cite{Zou}). While conventional designs employed wireless resource allocations for optimizing the security performance, this paper additionally exploits the UAVs' controllable mobility via trajectory design for further performance improvement. Also note that this paper is different from the interesting parallel work \cite{Lee} on the UAV-aided secrecy communications. In particular, \cite{Lee} considered the deterministic secrecy rate maximization by assuming perfect knowledge of eavesdropper's location at UAVs, while this paper studies the {\it worst-case} secrecy rate maximization by considering the uncertainty of such information. Therefore, the optimization approaches and insights are significantly different from those in \cite{Lee}.

\section{System Model}
We consider that a UAV transmitter (UAV $1$) delivers confidential information to a GN and the other UAV jammer (UAV $2$) cooperatively sends AN to confuse an on-ground suspicious eavesdropper.
Without loss of generality, we consider a three-dimensional (3D) Cartesian coordinate system, in which the GN and the eavesdropper are located on the ground with altitude zero, and horizontal locations $\mv{w}_0$ and $\mv{w}_e \in  \mathbb{R}^{2\times 1}$, respectively.
It is assumed that the two UAVs perfectly know the GN location $\mv{w}_0$ via proper information exchange, and partially know the eavesdropper location $\mv{w}_e$ by e.g. monitoring its potential information leakage \cite{Mukherjee2012}.
More specifically, let $\mv{\tilde{w}}_e$ denote the estimated eavesdropper location by the two UAVs.
We consider a bounded eavesdropper location error model with $\epsilon$ denoting the maximum estimation error that is known by the UAVs. Accordingly, we have $\!\mv{{w}}_e \!\in\!
\Theta\!\triangleq\!\{\!||\mv{\tilde{w}}_e\!-\!\mv{w}_e||\!\leq\!\epsilon\!\}$,
where $||\!\cdot\!||$ denotes the Euclidean norm.

We focus on a particular UAV mission/flight period with duration $T$, which is discretized into $N$ time slots with equal duration $t_s\!\!=\!\!T/N$.
The two UAVs fly at two given altitudes $H_1$ and $H_2$, respectively, which are pre-determined properly to avoid their collision.\footnote{Our results are extendible to the case when the two UAVs fly at the same altitude, by taking into account the minimum inter-UAV distance constraints for collision avoidance.} For each UAV $i\!\in\!\{1,2\}$, let ${\mv{q}}_i[n]$ denote its horizontal location at slot $n\!\in\!\{1,\ldots,N\}$, and $\mv{q}_i[0]$ and $\mv{q}_i[N+1]$ denote its initial and final locations, respectively. Let $\tilde{V}_i$ denote the maximum speed of UAV $i\!\in\!\{1,2\}$, and $V_i\!=\!\tilde{V}_i t_s$ denote the maximum displacement of that UAV between two consecutive slots.
It then follows that
\begin{align}\label{trajectory}
||\mv{q}_{i}[n\!+\!1]\!-\!\mv{q}_{i}[n]||\leq V_{i}, \forall i\in\{1,2\}, n\!\in\!\{0,\ldots,N\}.
\end{align}

In practice, the air-to-ground wireless channels are dominated by the LoS link.
As commonly adopted in the UAV communication literature \cite{zhangguangchi,cui,li,qian}, we adopt the free-space wireless channel model, in which the channel power gain from UAV $i$ to the GN at slot $n$ is given by
\begin{align}
g_i[n]=\frac{\beta_0}{d_{i,0}^2[n]}=\frac{\beta_0}{||\mv{q}_i[n]-\mv{w}_0||^2 + H_i^2},
\end{align}
where $\beta_0$ denotes the channel power gain at the reference distance of $1$ m and $d_{i,0}[n]=\sqrt{||\mv{q}_i[n]-\mv{w}_0||^2 + H_i^2}$ is the distance from UAV $i$ to the GN at slot $n$.
Similarly, the channel power gain from UAV $i$ to the eavesdropper at slot $n$ is
\begin{align}
h_{i}[n]=\frac{\beta_0}{d_{i,e}^2[n]}=\frac{\beta_0}{||\mv{q}_i[n]-\mv{w}_e||^2 + H_i^2},
\end{align}
where $d_{i,e}[n] = \sqrt{||\mv{q}_i[n]-\mv{w}_e||^2 + H_i^2}$ is the distance from UAV $i$ to the eavesdropper at slot $n$.

Next, we consider the cooperative jamming between the two UAVs.
At time slot $n$, let $p_{1}[n]$ and $p_{2}[n]$ denote the transmit power by UAV $1$ for communication and UAV $2$ for cooperative jamming, respectively.
Then, the achievable rates from UAV $1$ to the GN and the eavesdropper (in bps/Hz) at slot $n$ are respectively given as
\begin{align}\label{R_UG}
r_{0}[n]
=\log_2\left(1+\frac{g_{1}[n]p_1[n]}{g_{2}[n]p_{2}[n]+\sigma^2}\right),\\\label{R_UE}
r_{e}[n]=\log_2\left(1+\frac{h_{1}[n]p_1[n]}{h_{2}[n]p_{2}[n]+\sigma^2}\right).
\end{align}
Notice that each UAV only partially knows the eavesdropper location $\mv{w}_e$.
We are interested in the worst-case secrecy rate from UAV $1$ to the GN for each slot $n$, which is given by
\begin{align}\label{R_sec}
R[n]=\big[r_{0}[n]-\max\limits_{\mv{w}_e\in \Theta}r_{e}[n]\big]^+,
\end{align}
\!\!where $\![x]^+\!\!\triangleq\!\text{max}(x,\!0)$.
Furthermore, suppose that each UAV $\!i\!$ is subject to a maximum average power $\!P_\text{ave}\!$ and a maximum peak power $\!P_\text{peak}$.
Then we have
\begin{subequations}\label{power}
\begin{align}
&\frac{1}{N}\sum_{n=1}^{N} p_{i}[n] \le P_\text{ave}, \forall i\in\{1,2\}, \\
&p_{i}[n] \le P_\text{peak}, \forall i\in\{1,2\},n\in\{1,\ldots,N\}.
\end{align}
\end{subequations}

Our objective is to maximize the average secrecy rate from UAV $1$ to the GN over the whole period, by jointly optimizing the UAV trajectory $\{\mv{q}_i[n]\}$ and the transmit power allocation $\{p_i[n]\}$, subject to the trajectory constraints in (\ref{trajectory}) and the power constraints in (\ref{power}).
The problem of interest is formally formulated as
\begin{align}\label{PRO}
(\text{P1}):\max \limits_{\{\mv{q}_i[n],p_i[n]\}}\frac{1}{N}\sum_{n=1}^{N}R[n],
~\text{s.t.}~ (\ref{trajectory})~\text{and}~(\ref{power}).
\end{align}
Notice that the objective function in problem (P1) is a non-smooth (due to the operator $[\cdot]^+$) and non-concave function involving a ``$\max$'' operation (i.e., $\max\limits_{\mv{w}_e \in \Theta} r_{e}[n]$). Therefore, problem (P1) is a non-convex optimization problem that is difficult to be optimally solved in general.

\section{Proposed Solution to Problem (P1)}
In this section, we propose an efficient solution to problem (P1). First, to facilitate the derivation, we approximate $\!\!\max\limits_{\mv{w}_e \in \Theta} \!r_{e}[n]\!\!$ in the objective function as an explicit function. Notice that we have $\!\!\max\limits_{\mv{w}_e \in \Theta} \!h_1[n] \!\!=\!\! \tilde{h}_1[n] \!\!=\!\! \frac{\beta_0}{\left({||\mv{q}_1[n]-\mv{\tilde{w}}_e||}-\epsilon\right)^{2}+H_1^2}$, and $\!\min\limits_{\mv{w}_e \in \Theta} \!h_2[n] \!\!=\!\! \tilde{h}_2[n] \!\!=\!\! \frac{\beta_0}{\left({||\mv{q}_{2}[n]-\mv{\tilde{w}}_e||}+\epsilon\right)^2+H_2^2}$, where $\tilde{h}_1[n]$ corresponds to the best channel power gain from UAV $1$ to the eavesdropper that is attained at $\mv w_e[n]\!=\! \mv{\tilde{w}}_e+\frac{\mv{q}_1[n]-\mv{\tilde{w}}_e}{||\mv{q}_1[n]-\mv{\tilde{w}}_e||}\epsilon$, and
$\tilde{h}_{2}[n]$ corresponds to the worst channel power gain from UAV $2$ to the eavesdropper that is attained at $\!\mv w_e[n]\!\!=\!\! \mv{\tilde{w}}_e\!-\!\frac{\mv{q}_2[n]-\mv{\tilde{w}}_e}{||\mv{q}_2[n]-\mv{\tilde{w}}_e||}\epsilon$. Therefore, it is evident that
\begin{align}\label{r_e_approx}
\max_{\mv{w}_e\in\Theta} r_e[n] \le \tilde{r}_{e}[n]=\log_2\left(1+\frac{\tilde{h}_{1}[n]p_1[n]}{\tilde{h}_{2}[n]p_{2}[n]+\sigma^2}\right),
\end{align}
where $\tilde r_e[n]$ corresponds to an upper bound of the achievable rate from UAV $1$ to the eavesdropper.
By replacing $\max\limits_{\mv{w}_e\in\Theta} r_e[n]$ as $\tilde{r}_{e}[n]$ in (\ref{r_e_approx}), we obtain a lower bound of the secrecy rate of UAV $1$ at slot $n$ as
$\tilde R[n] = \left[r_{0}[n] - \tilde r_{e}[n]\right]^+$. Therefore, instead of directly maximizing $\frac{1}{N}\sum_{n=1}^N R[n]$ in (P1), we propose to maximize the explicit and mathematically tractable function $\frac{1}{N}\sum_{n=1}^N \tilde R[n]$ to obtain a lower bound for UAV $1$'s secrecy rate.
Furthermore, it is worth noting that, according to Lemma 1 in \cite{zhangguangchi}, the optimization with transmit power allocation always leads to a non-negative secrecy rate at each time slot. Therefore, we can equivalently omit the $[\cdot]^+$ operator in the objective function for the secrecy rate maximization.
As such, we approximate problem (P1) as the following problem:
\begin{align}\label{PRO2}
(\text{P2}):\max \limits_{\{\mv{q}_i[n],p_i[n]\}}&\frac{1}{N}\sum_{n=1}^{N}\bar{R}[n],~\text{s.t.}~(\ref{trajectory})~\text{and}~(\ref{power}),
\end{align}
where $\bar{R}[n]=r_{0}[n]-\tilde{r}_{e}[n], n\in\{1,...,N\}$.

Next, we focus on solving the approximate problem (P2), which, however, is still non-convex. To tackle this issue, we use the alternating optimization method to optimize the transmit power allocation $\{p_i[n]\}$ and UAV trajectories $\{\mv{q}_i[n]\}$ in an alternating manner, by considering the other to be given.

\subsection{Transmit Power Allocation}
First, we optimize the transmit power allocation $\{p_i[n]\}$ under given UAV trajectories $\{\mv{q}_i[n]\}$, for which the optimization problem is  expressed as
\begin{align}\label{sub_1} \max\limits_{\{p_i[n]\}}\frac{1}{N}\sum\limits_{n=1}^{N}\bar{R}[n],~~~ \text{s.t.}~~(\ref{power}),
\end{align}
where $\!\bar{R}[n]\!$ can be expressed in a concave-minus-concave form as

\begin{small}
\begin{align}\label{pro41}
\bar{R}&[n]= \log_2\left(\sigma^2+g_1[n]p_1[n]+g_{2}[n]p_{2}[n]\right)&\nonumber\\
&+\log_2\left(\sigma^2+\tilde{h}_{2}[n]p_{2}[n]\right)-\log_2\left(\sigma^2+g_{2}[n]p_{2}[n]\right)\nonumber\\&
-\log_2\left(\sigma^2+\tilde{h}_1[n]p_{1}[n]+\tilde{h}_{2}[n]p_{2}[n]\right).
\end{align}
\end{small}
As $\bar R[n]$ is a non-concave function with respect to $\{p_i[n]\}$, problem (\ref{sub_1}) is still a non-convex optimization problem. To tackle this issue, we apply the SCA technique to obtain a converged solution in an iterative manner. At each iteration $m \ge 1$, suppose that the local transmit power point is given as $\{p_i^{(m)}[n]\}$. Then we approximate $\bar{R}[n]$ as a lower bound $\bar{R}^{(m)}[n]$ in the following by using the first-order Taylor expansion.

\begin{small}
\begin{align}\label{R1_app}
&\bar{R}[n]\geq \bar{R}^{(m)}[n]\triangleq
\log_2\left(\sigma^2+g_{1}[n]p_1[n]+g_{2}[n]p_{2}[n]\right)
\\&+\!\log_2\left(\sigma^2\!+\!\tilde{h}_{2}[n]p_{2}[n]\right)
\!\!-\!\!\frac{\tilde{h}_{1}[n]\left(p_1[n]-p^{(m)}_{1}[n]\right)}{\ln2\left(\sigma^2\!+\!\tilde{h}_{1}[n]p^{(m)}_{1}[n]\!+\!\tilde{h}_{2}[n]p^{(m)}_{2}[n]\right)}\nonumber
\\&\!-\!\frac{g_{2}[n]\left(p_{2}[n]\!-\!p^{(m)}_{2}[n]\right)}{\ln2\left(\sigma^2\!+\!g_{2}[n]p^{(m)}_{2}[n]\right)}\!-\!\frac{\tilde{h}_{2}[n]\left(p_{2}[n]-p^{(m)}_{2}[n]\right)}{\ln2\left(\sigma^2\!\!+\!\!\tilde{h}_{1}[n]p^{(m)}_{1}[n]\!\!+\!\!\tilde{h}_{2}[n]p^{(m)}_{2}[n]\right)}
\!\nonumber\\
&\!-\!\log_2\left(\sigma^2\!\!+\!\!g_{2}[n]p^{(m)}_{2}[n]\right)\!\!-\!\!\log_2\left(\sigma^2\!\!+\!\!\tilde{h}_1[n]p^{(m)}_{1}[n]\!\!+\!\!\tilde{h}_{2}[n]p^{(m)}_{2}[n]\right)
.\nonumber
\end{align}
\end{small}By replacing $\bar R[n]$ as $\bar R^{(m)}[n]$, problem (\ref{sub_1}) is approximated as a convex optimization problem that can thus be solved by standard convex optimization techniques such as CVX [11]. Then, at each iteration $(m+1)$, we set the local point $\{p_i^{(m+1)}[n]\}$ as the optimal solution to the approximate problem in the previous iteration $m$. As the iteration converges, we can obtain a converged solution to problem (\ref{sub_1}).

\subsection{UAV Trajectory Design}
Next, we optimize the UAV trajectory $\{\mv{q}_i[n]\}$ under given transmit power allocation $\{p_i[n]\}$, for which the problem is expressed as
\begin{align}\label{sub_2}
\max\limits_{\{\mv{q}_i[n]\}}~~\frac{1}{N}\sum\limits_{n=1}^{N}\bar{R}[n], ~~~\text{s.t.}~~ (\ref{trajectory}).
\end{align}

By introducing auxiliary variables $\{\zeta[n]\}$, $\{\xi[n]\}$ and $\{\tau[n]\}$, we re-express problem (\ref{sub_2}) equivalently as

\begin{small}
\begin{subequations}\label{sub_2_1}
\begin{align}
&\max\limits_{\{\mv{q}_i[n],\zeta[n],\xi[n],\tau[n]\}}~\frac{1}{N}\sum\limits_{n=1}^{N}
\hat{R}[n],\nonumber\\
&~~~\text{s.t.}~~\zeta[n]\!\leq\!||\mv{q}_2[n]-\mv{w}_0||^2+H_2^2, \forall n\in\{1,\ldots,N\},\label{non_con1}\\
&~~\!\xi[n]\!\leq \! \left({||\mv{q}_1[n]-\mv{\tilde{w}}_e||}\!-\!\epsilon\right)^2\!+\!H_1^2, \forall n\in\{1,\ldots,N\},\label{non_con3}\\
&~~\!\tau[n]\leq \left({||\mv{q}_2[n]-\mv{\tilde{w}}_e||}\!+\!\epsilon\right)^2\!+\!H_2^2, \forall n\in\{1,\ldots,N\},\label{non_con4}\\
&~~\zeta[n]\!\geq\! H_2^2,\xi[n]\!\geq \! H_1^2,\tau[n]\!\geq\! H_2^2\!+\!\epsilon^2, \forall n\in\{1,\ldots,N\},\label{t1_1}\\
&~~~~~~~~~(\ref{trajectory})\nonumber,
\end{align}
\end{subequations}
\end{small}
where
\begin{small}
\begin{align} \label{non_con6}
&\!\!\hat{R}[n]\!\!=\!\!
-\!\!\log_2\!\!\left(\!\sigma^2\!\!+\!\!\frac{\beta_0p_{2}[n]}{\zeta[n]}\!\right)
\!\!-\!\!\log_2\!\!\left(\!\sigma^2\!\!+\!\!
\frac{\beta_0p_{1}[n]}{\xi[n]}\!\!+\!\!\frac{\beta_0p_{2}[n]}
{\tau[n]}\!\right)\\
+&\log_2\left(\sigma^2\!+\!\tilde{h}_{2}
[n]p_{2}[n]\right)
\!+\!\log_2\left(\sigma^2\!+\!g_{1}[n]p_{1}[n]\!+\!g_{2}[n]p_{2}[n]
\right)\!.\nonumber
\end{align}
\end{small}
\!\!Notice that the constraints in (\ref{non_con1}), (\ref{non_con3}), and (\ref{non_con4}) are non-convex, as their right-hand-side (RHS) terms are all convex. Also, $\hat R[n]$ in the objective function is non-concave, as the third and fourth terms are convex functions with respect to $\{\mv{q}_i[n]\}$. Therefore, problem (15) is non-convex.
To tackle the non-convexity, we apply the SCA technique to solve problem (15) in an iterative way.
Given a local UAV trajectory point $\{\mv{q}^{(m)}_i[n]\}$ at iteration $m\ge 1$, we have the approximate constraints and objective functions as follows based on the first-order Taylor expansion.

\begin{small}
\begin{subequations}
\begin{align}
&\zeta[n]\!-\!||\mv{q}^{(m)}_{2}[n]-\mv{w}_0||^2\!-\!2(\mv{q}^{(m)}_{2}[n]-\mv{w}_0)^T\!\!\times\!\left(\mv{q}_2[n]-\mv{q}^{(m)}_{2}[n]\right)\nonumber\\
&~~~~~~~~~~\leq H_2^2,\forall n\in\{1,\ldots,N\},\label{Ta_1}\\
&\nonumber \xi[n]\!-\!||\mv{q}_{1}^{(m)}[n]-\mv{\tilde{w}}_e||^2\!-\!2(\mv{q}_{1}^{(m)}[n]-\mv{\tilde{w}}_e)^T\!\!\times\!\left(\mv{q}_1[n]-\mv{q}_{1}^{(m)}[n]\right)\!\\ &~~~~~~~~~~\leq\!-2\epsilon{||\mv{q}_1[n]\!-\!\mv{\tilde{w}}_e||}\!+\!\epsilon^2+H_1^2,\forall n\in\{1,\ldots,N\},\label{Ta_2}\\
&\nonumber\tau[n]\!\!-\!\!\epsilon^2\!\!\!-\!\!H_2^2\!\!-\!\!||\mv{q}_{2}^{(m)}[n]\!\!-\!\!\mv{\tilde{w}}_e||^2\!\!\!-\!\!2\bigg(\!\mv{q}_{2}^{(m)}[n]\!\!-\!\!\mv{\tilde{w}}_e\!+\!\!\frac{\mv{q}_{2}^{(m)}[n]\!-\!\mv{\tilde{w}}_e}{||\mv{q}_{2}^{(m)}[n]\!-\!\mv{\tilde{w}}_e||}\epsilon\bigg)^T\\ &\!\!\times\!\!\left(\mv{q}_2[n]\!-\!\mv{q}_{2}^{(m)}[n]\right)\!\!\leq\!\!2\epsilon{||\mv{q}_{2}^{(m)}[n]\!-\!\mv{\tilde{w}}_e||},\forall n\in\{1,\ldots,N\},\label{Ta_3}\\
&\hat{R}[n]\!\geq \!\hat{R}^{(m)}[n]\!\triangleq
\!\!-\!\!\log_2\!\!\left(\!\sigma^2\!\!+\!\!\frac{\beta_0p_{2}[n]}{\zeta[n]}\!\right)
\!\!-\!\!\log_2\!\!\left(\!\sigma^2\!\!+\!\!
\frac{\beta_0p_{1}[n]}{\xi[n]}\!\!+\!\!\frac{\beta_0p_{2}[n]}
{\tau[n]}\!\right)\nonumber\\\nonumber
&\!+\!\!a\!\left(\!||\mv{q}_{2}\![n]\!\!-\!\!\mv{w}_0||^2\!\!\!\!-\!||\mv{q}_{2}^{(\!m\!)}\![n]\!\!\!-\!\!\mv{w}_0\!||^2\!\right)\!\!+\!\!
b\!\left(\!||\mv{q}_1\![n]\!\!-\!\!\mv{w}_0\!||^2\!\!\!-\!\!||\mv{q}_{1}^{(\!m\!)}\![n]\!\!-\!\!\mv{w}_0||^2\!\right)\\
&\!+\!\!c\left(\left({||\mv{q}_2[n]\!\!-\!\!\mv{\tilde{w}}_e||}\!\!+\!\epsilon\right)^2\!\!\!\!-\!\!\left({||\mv{q}_{2}^{(m)}[n]\!\!-\!\!\mv{\tilde{w}}_e||}\!+\!\!\epsilon\right)^2\right)+d,
\end{align}
\end{subequations}
\end{small}
where $a\!=\!-\frac{g_{2}^2[n]p_{2}[n]}{\ln2\beta_0\left(\sigma^2
+g_1[n]p_{1}[n]\!+\!g_{2}[n]p_{2}[n]\right)}, ~ b\!=\!-\frac{g_{1}^2[n]p_{1}[n]}{\ln2\beta_0\left(\sigma^2
+g_{1}[n]p_{1}[n]\!+\!g_{2}[n]p_{2}[n]
\right)}, c\!=\!-\frac{\tilde{h}_{2}^2[n]p_{2}[n]}{\ln2\beta_0\left(\sigma^2
+\tilde{h}_{2}[n]p_{2}[n]\right)},$ and $d\!=\!\log_2\left(\sigma^2\!+ \frac{\beta_0p_{2}[n]}{\left({||\mv{q}^{(m)}_{2}[n]-\mv{\tilde{w}}_e||}+\epsilon\right)^2+H_2^2}\right)\!+\!\log_2\bigg(\sigma^2\!\!+\\\!\!\frac{\beta_0p_1[n]}{||\mv{q}^{(m)}_1[n]-\mv{w}_0||^2 + H_i^2}\!\!+\!\!\frac{\beta_0p_2[n]}{||\mv{q}^{(m)}_2[n]-\mv{w}_0||^2 + H_i^2}
\bigg)$.
Replacing $\!\hat{R}[n]$, (\ref{non_con1}), (\ref{non_con3}), and (\ref{non_con4}) with $\!\hat{R}^{(m)}[n]$, (\ref{Ta_1}), (\ref{Ta_2}), and (\ref{Ta_3}), respectively, problem (\ref{sub_2_1}) is equivalently expressed as the following convex optimization problem that can be efficiently solved by CVX.
\begin{align}\label{pro5}
&\max\limits_{\{\mv{q}_i[n],\zeta[n],\xi[n],\tau[n]\}}~\frac{1}{N}\sum_{n=1}^N
\hat{R}^{(m)}[n],
\\
&~~~~~~~~~~\text{s.t.}~ (\ref{Ta_1}),(\ref{Ta_2}),(\ref{Ta_3}),
(\ref{t1_1})~\text{and}~(\ref{trajectory}).\nonumber
\end{align}
 By iteratively updating the local UAV trajectory point $\{\mv{q}^{(m)}_i[n]\}$ at the $(m+1)$-th iteration as the optimal solution to the approximate problem in the previous iteration $m$, we can obtain a converged solution to problem (\ref{sub_2}).

To sum up, we solve for the transmit power $\{p_i[n]\}$ and the trajectory $\{\mv{q}_i[n]\}$ in an alternating manner above, and accordingly, we obtain an efficient solution to problem (P2).
As the objective value of problem (P2) is monotonically non-decreasing after each iteration and the objective value of problem (P2) is finite, the proposed alternating optimization based algorithm is guaranteed to converge \cite{chanllenges,wuqingqing}. In Section IV, we will conduct simulations to show the effectiveness of the proposed algorithm.

\section{Numerical Results}
In this section, we provide simulation results to verify the performance of our proposed design.
For comparison, we consider a heuristic UAV trajectory design, namely the fly-hover-fly trajectory.
In this design, each of the two UAVs first flies straightly at the maximum speed from the initial location to the respective hovering location (above the GN to help communication for UAV $1$ and above the eavesdropper to facilitate jamming for UAV $2$, respectively), then hovers with the maximum duration, and finally flies straightly at their maximum speed to the final location.
Based on the fly-hover-fly trajectory, we consider both constant power allocation (i.e., $p_i[n] = P_\text{ave}, \forall i,n$) and adaptive power allocation (i.e., by solving problem (\ref{sub_1}) under given trajectory). Furthermore, for our proposed design, we use the fly-hover-fly trajectory with constant power allocation as the initial point for optimization.

In the simulation, we set $\mv{w}_0\!\! =\!\! (0,0)$, $\mv{\tilde{w}}_e\!\! =\!\! (200~\textrm{m},0)$, $H_1 \!\! =\!\!100~\textrm{m}$, $H_2 \!\! =\!\!110~\textrm{m}$, $V_{1}\!\! =\!\!V_{2}\!\! =\!\!10~\textrm{m/s}$, $P_\text{ave}\!\! =\!\!30~\textrm{dBm}$, $P_\text{peak}\!\! =\!\!4P_\text{ave}$, $\epsilon\!\! =\!\!10~\textrm{m}$, $\beta_0/ \sigma^2 \!\! =\!\! 80~\textrm{dB}$, $\mv{q}_1[0]\!\! = \mv{q}_2[0]\!\! =\!\!(100~\textrm{m},500~\textrm{m})$, and $\mv{q}_1[N+1]\!\! =\!\! \mv{q}_2[N+1]\!\! =\!\!(100~\textrm{m},-500~\textrm{m})$.

\begin{figure}[!t]
\centering
\includegraphics[width=6cm]{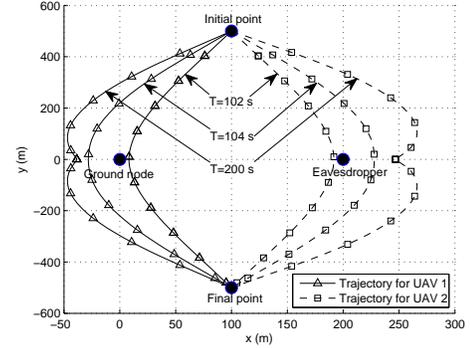}
\caption{Obtained UAV trajectories by the proposed design.}
\label{Fig_traj}
\end{figure}
Fig. \ref{Fig_traj} shows the obtained trajectories of two UAVs by our proposed design, under mission duration $T\!\! =\!\!102$ s, $T\!\! =\!\!104$ s, and $T\!\! =\!\!200$ s, respectively. It is observed that when $T$ is large (e.g., $T\!\! =\!\!200$ s), the two UAVs first fly to the hovering locations (that are close to but slightly away from the GN or the eavesdropper, respectively) following arc paths, then hover there with longest duration, and finally fly to the final locations in symmetric arc paths. Intuitively, staying at the corresponding hovering location, UAV 1 can achieve the best secrecy communication performance by effectively balancing between the desirable information transmission to the GN versus undesirable information leakage to the eavesdropper, while UAV 2 can achieve the best cooperative jamming performance by balancing between the desirable jamming towards the eavesdropper versus the undesirable interference with the GN. By contrast, it is also observed that when $T$ becomes $T\!\! =\!\!102$ s or $104$ s, the two UAVs fly at the maximum speed towards the hovering locations as close as possible, but they cannot exactly reach there due to the time and speed limitations. It is expected that as $T$ increase, the UAVs can fly closer towards their respective hovering locations and hover there with longer durations, thus leading to higher average achievable secrecy rates, as illustrated next.

\begin{figure}[!t]
\centering
\includegraphics[width=6cm]{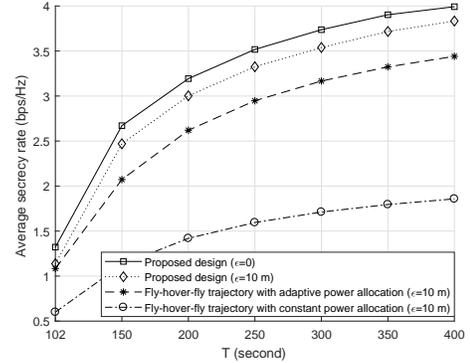}
\caption{Average secrecy rate versus mission duration $T$.}
\label{Fig_rate}
\end{figure}
Fig. \ref{Fig_rate} shows the average achievable secrecy rate versus mission duration $T$. For comparison, we also consider the case with perfect eavesdropper location information, i.e., $\epsilon\!\!=\!\!0$ m. It is observed that as $T$ increases, the secrecy rates achieved by all the four schemes increase, as the two UAVs can stay longer at the corresponding hovering locations for better communication/jamming performance as explained above. When $T$ is small (e.g. $T\!\!=\!\!102$ s), our proposed design is observed to have a similar performance as the fly-hover-fly trajectory with adaptive power allocation. This is due to the fact that in this case, the block duration is only sufficient for the two UAVs to fly from the initial to final locations, and there is no additional time to adjust the trajectory for communication performance optimization. When $T$ becomes large, the proposed design is observed to significantly outperform the two benchmark schemes with fly-hover-fly trajectory. Furthermore, it is observed that there is only a slight performance gap between the proposed design with perfect eavesdropper location information ($\epsilon\!\!=\!\!0$ m) versus that with imperfect eavesdropper location information ($\epsilon\!\!=\!\!10$ m). This shows the effectiveness of our proposed approximation.

\section{Conclusion}
In this letter, we presented a new cooperative jamming approach to secure the UAV communication, by enabling another nearby UAV to cooperatively jam the potential eavesdropper on the ground.
By exploiting the UAVs' controllable mobility, we maximized the average secrecy rate by optimizing the UAVs' trajectories, jointly with their transmit power allocations.
How to extend the results to other scenarios, e.g., with multiple UAVs and multiple antennas, are interesting directions worth further investigation.

\end{document}